\documentclass[aps,prl,twocolumn,showpacs,superscriptaddress,groupedaddress]{revtex4}  
\usepackage{graphicx,epsfig}  
\usepackage{dcolumn}   
\usepackage{bm}        
\usepackage{amssymb}   
\usepackage{color}
\usepackage[caption = false]{subfig}
\begin{document}

\title{Experimental observation of precursor solitons in a flowing complex plasma}

\author{Surabhi Jaiswal}
\email{surabhijaiswal73@gmail.com}
\affiliation{Institute for Plasma Research, Bhat, Gandhinagar-382428, India.}
\author{P. Bandyopadhyay}
\affiliation{Institute for Plasma Research, Bhat, Gandhinagar-382428, India.}

\author{A. Sen}
\affiliation{Institute for Plasma Research, Bhat, Gandhinagar-382428, India.}

\begin{abstract}
The excitation of precursor solitons ahead of a rapidly moving object in a fluid, a spectacular phenomenon in hydrodynamics that has often been observed ahead of moving ships, has surprisingly not been investigated in 
plasmas where the fluid model holds good for low frequency excitations such as ion acoustic waves. In this paper 
we report the first experimental observation of precursor solitons in a flowing dusty plasma. The nonlinear solitary dust acoustic waves (DAWs) are excited by a supersonic mass flow of the dust particles over an electrostatic potential hill. In a frame where the fluid is stationary and the hill is moving the solitons propagate in the upstream direction as precursors while wake structures consisting of linear DAWs are seen to propagate in the downstream region.  A theoretical explanation of these excitations based on the forced Korteweg-deVries model equation is provided and their practical implications in situations involving a charged object moving in a plasma are discussed.  
\end{abstract}

\maketitle

Wave patterns generated by an object moving in a fluid have long been a classical topic of research both for its fundamental significance as well as for its numerous practical implications in oceanography, atmospheric dynamics and various other engineering applications \cite{Stoker1992,Holthuijsen2007,Huang2009,Hermans2011}. A well known phenomenon is that of wakefields generated behind a moving object such as a boat traveling on a lake surface.
An important question, that has received much attention in hydrodynamics, is what happens when the speed of the moving object approaches and crosses the phase speed of the characteristic linear mode of the medium. Theoretical and numerical studies \cite{Akylas1984,Ertekin1986,Smythe1990,ZHANG2001} have shown that for speeds above this critical velocity the object generates not only wake patterns behind it 
but also radiates a steady stream of solitons ahead of it in the upstream region that move away from it at a faster speed. This fascinating phenomenon that has also been observed experimentally (using ship models moving in shallow channels) has been successfully modeled using nonlinear evolution equations like the forced Korteweg deVries (fKdV) equation or the forced Boussinesque equation \cite{Zabusky1971,Wu1987,Lee1989,Jones1999,Binder2014}. These precursor solitons are also seen to be excited if instead of moving the object the fluid is made to flow at a supercritical speed over a stationary object resting at the bottom of the fluid channel \cite{Wu1987, Binder2014}. \par
Wakefield patterns also occur behind a charged object (particle) moving in a plasma medium 
 \cite{Gurevich1980,Havnes1996,Brattli2002a,Engwall2006a,Sylla2011,Hutchinson2011,Block2012}. An open and interesting question to ask is whether the phenomenon of precursor wave excitations can occur in a plasma? In a recent theoretical investigation \cite{Sen2015} a model calculation predicts the excitation of precursor solitons in a plasma due to the passage of a charged object travelling faster than the ion acoustic velocity. However, to the best of our knowledge, there has so far been no experimental observation or demonstration of {\it precursor} solitons in a plasma medium although standard dust acoustic solitons triggered by short pulse excitations have been studied in a number of experiments \cite{Nosenko2002, Samsonov2002, Bandyopadhyay2008, Heidemann2009, Zhdanov2010}. The existence of upstream solitonic excitations could have wide ranging applications and also open up new areas of fundamental research in flowing plasma dynamics. Hence it is important to establish the experimental feasibility of such a phenomenon.  In this paper we report the first experimental observation of precursor solitons in a flowing dusty plasma.  A dusty plasma consisting of electrons, ions and charged micron sized macroparticles behaves in a manner similar to a conventional plasma but has additional collective modes due to the dust dynamics such as the very low frequency dust acoustic wave (DAW). This mode can be visually identified by illuminating the dust particles with a laser and observing (or examining video images of) the movement of the compressed and rarefied portions of the dust component.  Such a non-perturbative means of detecting and measuring collective excitations is a major diagnostic convenience of dusty plasmas that has been well exploited in many past experiments \cite{Bandyopadhyay2008,Pieper1996,Samsonov1999,Pramanik2002} and has motivated our present choice of a dusty plasma medium for the study of precursor solitons. As a further experimental convenience, rather than moving a charged object in the plasma, we have chosen a configuration in which the dusty plasma flows over an electrostatic potential hill representing a stationary charged object. \par
\begin{figure}[!hb]
\includegraphics[scale=0.45]{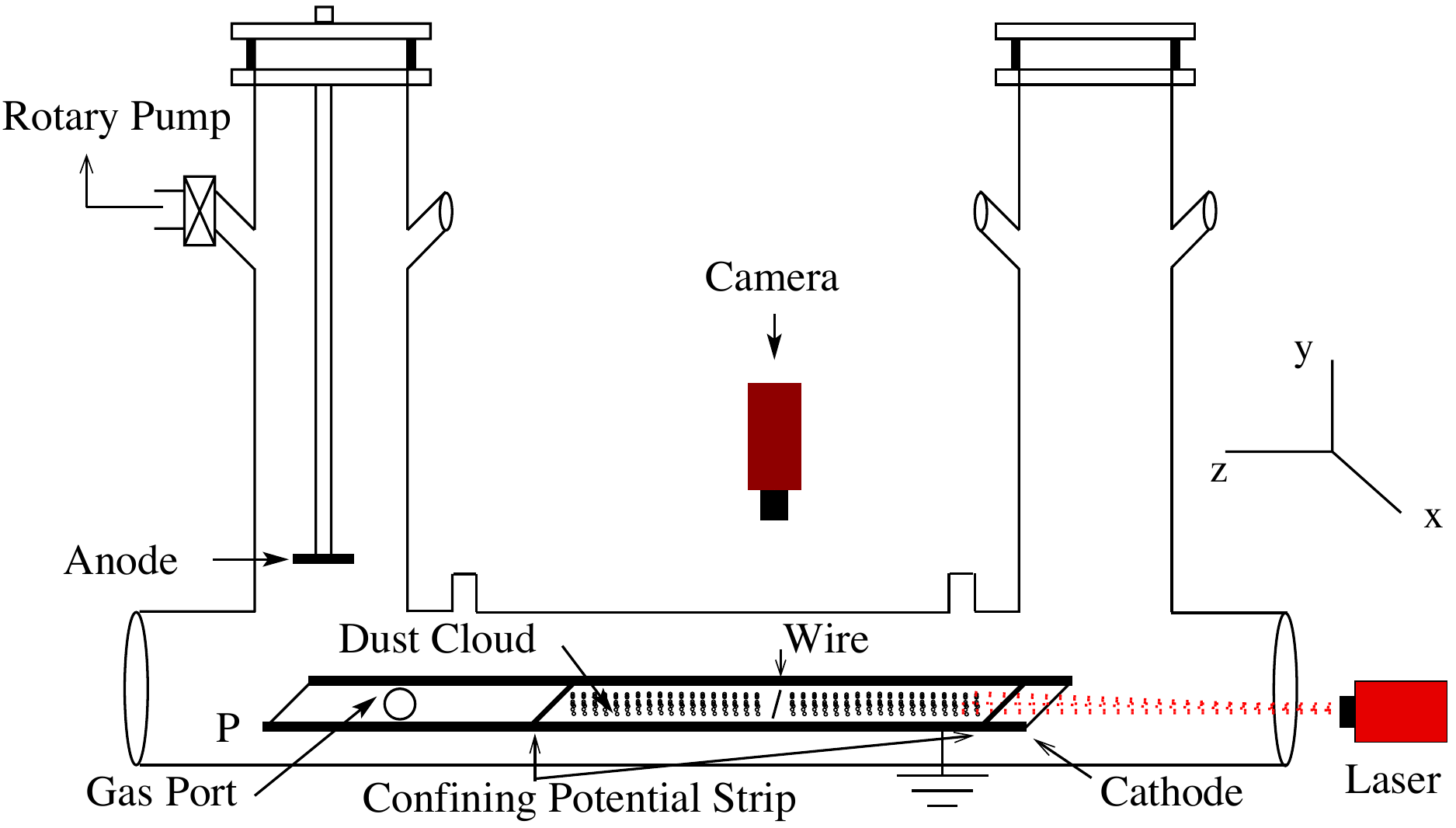}
\caption{\label{fig:setup} A schematic of the Dusty Plasma Experimental (DPEx) Device}
\end{figure}
\begin{figure}[!ht]
\includegraphics[scale=0.7]{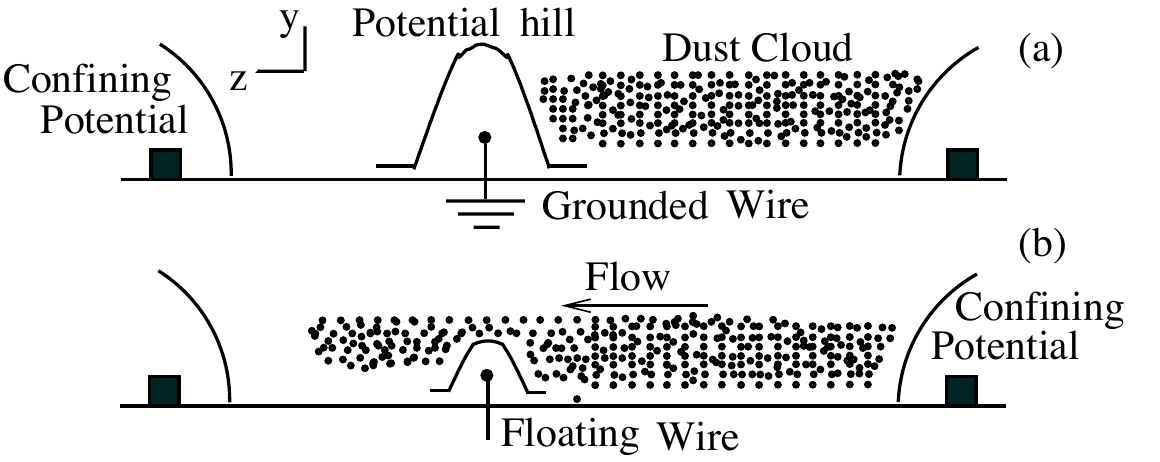}
\caption{\label{fig:potential} (a) Equilibrium dust configuration with the potential hill created by a grounded wire and b) dust flow induced by sudden lowering of the potential hill by a floating wire. }
\end{figure}

The experiments have been carried out in a specially designed Dusty Plasma Experimental (DPEx) device (shown in Fig.~\ref{fig:setup}) that is constructed using a $\Pi$-shaped glass chamber with a configuration similar to the PK-4 device \cite{Usachev2004} and with a facility to make the dusty plasma flow in a controlled manner. For a complete description of the device the reader is directed to reference \cite{Jaiswal2015}. A direct current (DC) glow discharge Argon plasma is created by applying a voltage of $~400~volts$ between a stainless steel (SS) disc anode and a cathode consisting of a long grounded tray. Micron sized kaolin particles (with a size dispersion ranging from $4$ to $6$ microns) sprinkled on the cathode plate get negatively charged and levitate to constitute the dust component. Two SS strips placed $25~cms$ apart on the cathode provide confinement of the dust particles in the axial direction through their sheath electric fields. A copper wire mounted midway between the SS strips, at a height of $1~cm$ from the cathode plate, that can be grounded or maintained at a chosen potential upto the floating potential, creates a potential hill that acts as an obstruction to the dusty plasma flow. The potential hill is used to initially to confine the dust particles between the right SS strip and the wire (see Fig.~\ref{fig:potential}(a)). 
The device is operated at a  working pressure of $P=0.090$ mbar and a discharge voltage of V$_a=350~volts$ at which the discharge current is $I_p \sim 2~ mA$.
Plasma parameters are measured using single Langmuir and emissive probes and typical experimental values are: plasma density $n_{i} \approx 1 \times 10^{14} m^{-3}$, electron temperature  $T_{e} \approx 5 eV$ and ion temperature $T_{i} \approx 0.03 eV$.  
A CCD camera captures the image of the dust cloud illuminated by a red diode laser.   From the video images of the cloud the dust density is estimated to be $n_{d} \approx 1 \times 10^{9} m^{-3}$.  
The average dust mass is $m_{d} \approx 1.7 \times 10^{-13} kg $  and the average charge on a dust particle (inferred from the plasma parameters and the particle size) is approximately $Q_{d} \approx 4.5 \times 10^{4} e$. The electron density is obtained from the quasi-neutrality condition by taking account of the dust contribution. Based on these parameter values the typical magnitude of the linear phase velocity of a DAW turns out to be $v_{ph} \approx 2.4 cm/sec$. Our estimated values of the dust charge and the phase velocity of DAW agree quite well with earlier investigations that were carried out in dusty plasmas with similar experimental parameters \cite{{Nakamura2012,Barkan1995}}. We have also independently measured the phase velocity by analyzing the propagation speed of spontaneously excited DAWs and obtained a value of $\approx 2.5 cm/sec$ which is very close to the theoretically estimated value. \par
\begin{figure}[!h]
\includegraphics[scale=0.55]{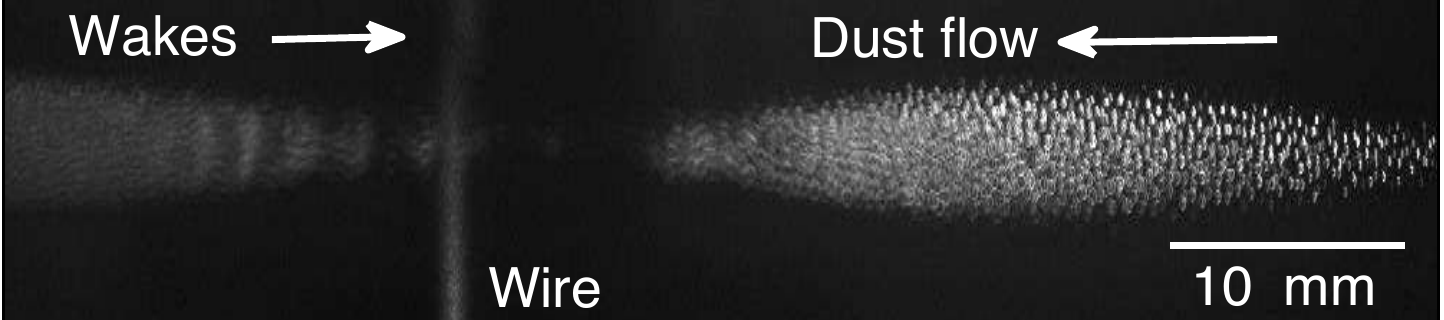}
\caption{\label{fig:only_wakes} Generation of wakes due to the subsonic flow of the dust fluid over the wire. }
\end{figure}

Our basic equilibrium, in the absence of the wire, consists of a long dust cloud confined between the two potential strips. This equilibrium is seen to occasionally develop spontaneous excitations \cite{Thoma2006} of dust acoustic waves (DAWs) particularly when the background pressure is below $\sim 0.07$ mbar. We therefore maintain our background pressure at $\sim0.09$ mbar and above to inhibit such spontaneous excitations.  We have also ascertained that the dust density does not influence the spontaneous excitation process by examining equilibria with different dust densities under the same discharge conditions and found no excitations even for low dust densities as long as the neutral pressure is maintained at $\sim0.09$ mbar and above. In the presence of the wire we start with an equilibrium of dust particles that are trapped between the right steel strip and the potential hill created by the wire. This trapped stationary dust cloud is then made to flow from right to left by suddenly reducing the height of the potential hill above the wire and thereby removing the barrier that obstructs them from getting to the left half of the cathode region. Fig.~\ref{fig:potential}(b) shows a schematic diagram of this situation when the height of the potential is reduced by removing the grounding from the wire and allowing it to have a floating potential. The height of the potential and hence the speed of flow of the particles, can be precisely regulated by drawing different amounts of currents from the wire using an external resistance. The flow speed of the dust cloud is determined by the amount of initial lowering of the height of the wire potential, the density of the dust particles and the magnitude of the neutral pressure in the chamber. In our experiments, for the given set of observations reported, we hold the dust density as well as the neutral pressure at constant values and change the flow speed by lowering the potential hill to different heights. Upon release from the potential barrier the flowing dust particles quickly attain a terminal velocity $U$ due to the slowing down effect from the neutral drag force \cite{Jaiswal2015}. This velocity is found to remain uniform over a substantial region of the device till the time that the cloud runs out of dust particles. The fluid velocity is measured by analyzing a few successive frames of the video image near the copper wire by using the Particle Image Velocimetry (PIV) tool in the Matlab software package \cite{Thielicke2014}.
Using the above technique we have made a number of experimental runs to study the flow induced excitations of the dusty plasma as it passes over the potential barrier.  For flows that are slow compared to the DAW phase speed (subsonic) wakefields are excited in the left side of the wire traveling in the direction opposite to the flow. In the frame of the fluid where the hill is moving from left to right these wakefields are in the downstream region. The actual velocity of the wake is given by $U - v_{wm}$ where $U$ is the flow velocity and $v_{wm}$ is the measured velocity of the wake fronts \cite{Wu1987}. A typical experimental image of a subsonic flow case is shown in Fig.~\ref{fig:only_wakes} for  $U \approx 1.8 cm/sec$ and $v_{wm} \approx -0.5 cm/sec$. These excitations thus travel at a speed of $U - v_{wm} \approx  2.30~cm/sec$ which is close to the phase velocity of the linear DAWs. 
\begin{figure}[!ht]
\includegraphics[scale=0.5]{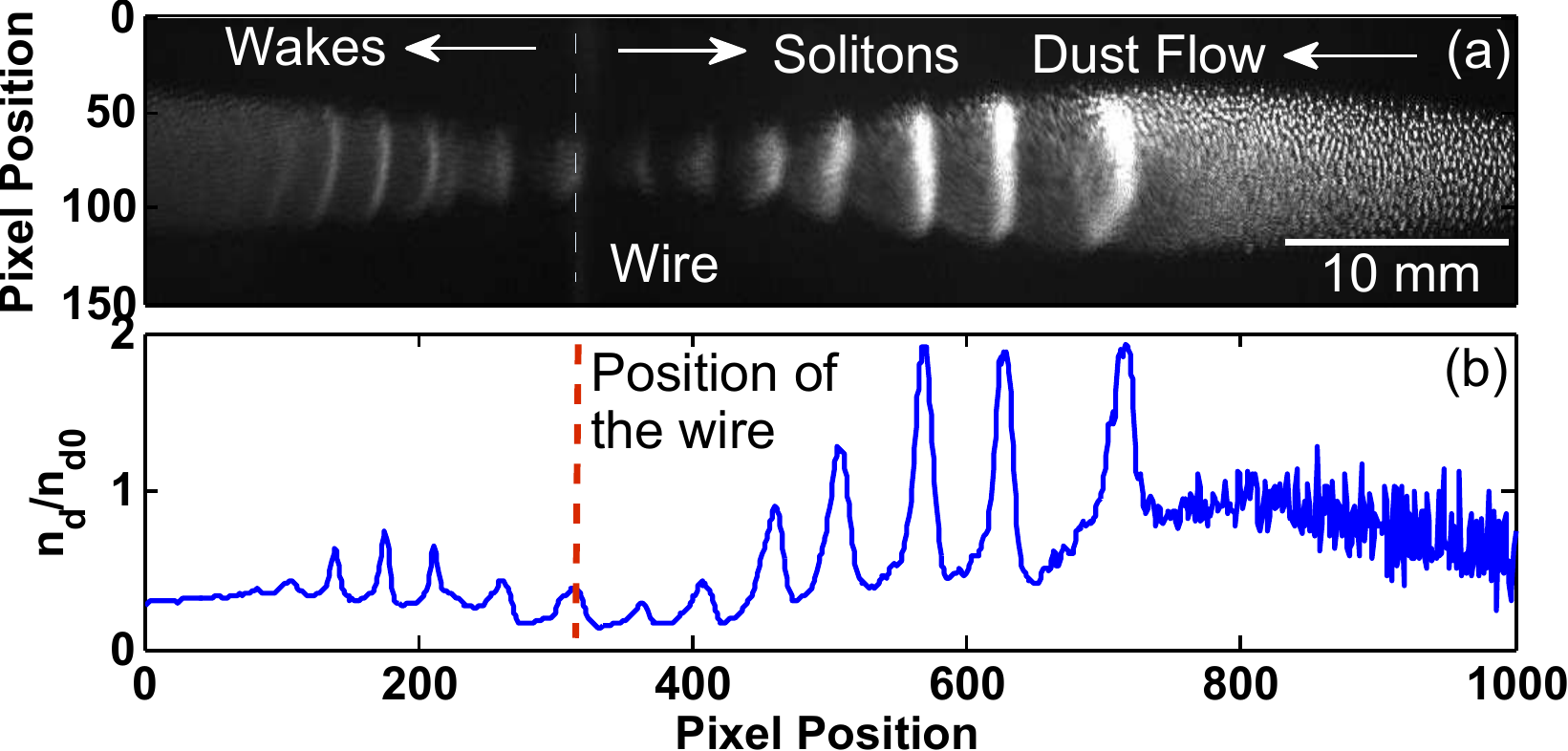}
\caption{\label{fig:real_image} (a) An experimental image of the excited precursor solitons and small amplitude wakes. (b) The intensity profile of Fig~\ref{fig:real_image}(a). }
\end{figure}
When the flow is made supersonic we notice a dramatic change in the nature of the excitations. In addition to the wake fields to the left of the wire we observe large solitary wave excitations to the right of the wire that travel in the direction opposite to the flow or in the frame of the fluid in the upstream direction of the moving potential hill. An experimental image of such excitations is shown in Fig.~\ref{fig:real_image} for $U \approx 2.65 cm/sec$. The measured velocity of these structures is $v_{sm} \approx -1.5~cms/sec$ and their actual speed after taking account of the flow is $v_{s} = U + v_{sm} \approx 4.15~cms/sec$ which is about 1.6 times the DAW speed.
Thus the speed of the flow (whether subsonic or supersonic) determines whether solitons are excited or not. Once the solitons are excited (for supersonic flow conditions) they continue to propagate at their characteristic nonlinear speed and the speed of the flow has no influence on their shape or propagation speed. The laboratory frame measurement of the speed of the soliton structure, $v_{sm}$, would however change if the flow speed changes in order to preserve the constancy of the soliton speed, namely, $v_{s}=U + v_{sm}$. As can be seen from Fig.~\ref{fig:real_image}, the space interval between successive solitons is nearly constant indicating that $U$ is uniform over the region of propagation of the solitons.  Note that since the flow velocity in this case is larger than the DAW speed the wake structures to the left of the wire are carried forward in the direction of the flow. These precursor pulses are emitted in regular time intervals and keep growing in amplitude as they travel to the right till they attain a saturated amplitude value. These saturated pulses travel faster than the dust acoustic speed. Their speed also depends on the size of their amplitude - a property typical of KdV solitons. These observations are consistent with theoretical predictions based on the forced KdV equation that has been derived for a charged object moving through a plasma medium \cite{Sen2015}. For a dusty plasma medium we have derived a corresponding forced KdV equation which is of the form, 
\begin{equation}
\frac{\partial n_{d1}}{\partial t} +A n_{d1} \frac{\partial n_{d1}}{\partial \xi} + \frac{1}{2}\frac{\partial^3 n_{d1}}{\partial \xi^3} = \frac{1}{2}\frac{\partial S_2}{\partial \xi}
\label{fKdv}
\end{equation} 
\noindent
where $n_{d1}$ is the perturbed dust density normalized to the equilibrium dust density and $\xi=(z - u_{ph} t)$ is the coordinate in the wave frame moving at phase velocity $u_{ph}$ normalized to the dust acoustic speed. 
The spatial coordinate $z$ is normalized by the Debye length ($\lambda_d$) and time is made dimensionless by the dust plasma  frequency ($\omega_{pd} $). The left hand side of (\ref{fKdv}) is the KdV equation describing the evolution of weakly nonlinear dispersive waves in a dusty plasma with the coefficient $A = \left[\delta^2 + (3\delta+\sigma_i)\sigma_i 
+ \frac{1}{2}\delta(1+\sigma_i^2)\right]/(\delta-1)^2$.
\noindent
\begin{figure}%
\includegraphics[scale=0.65]{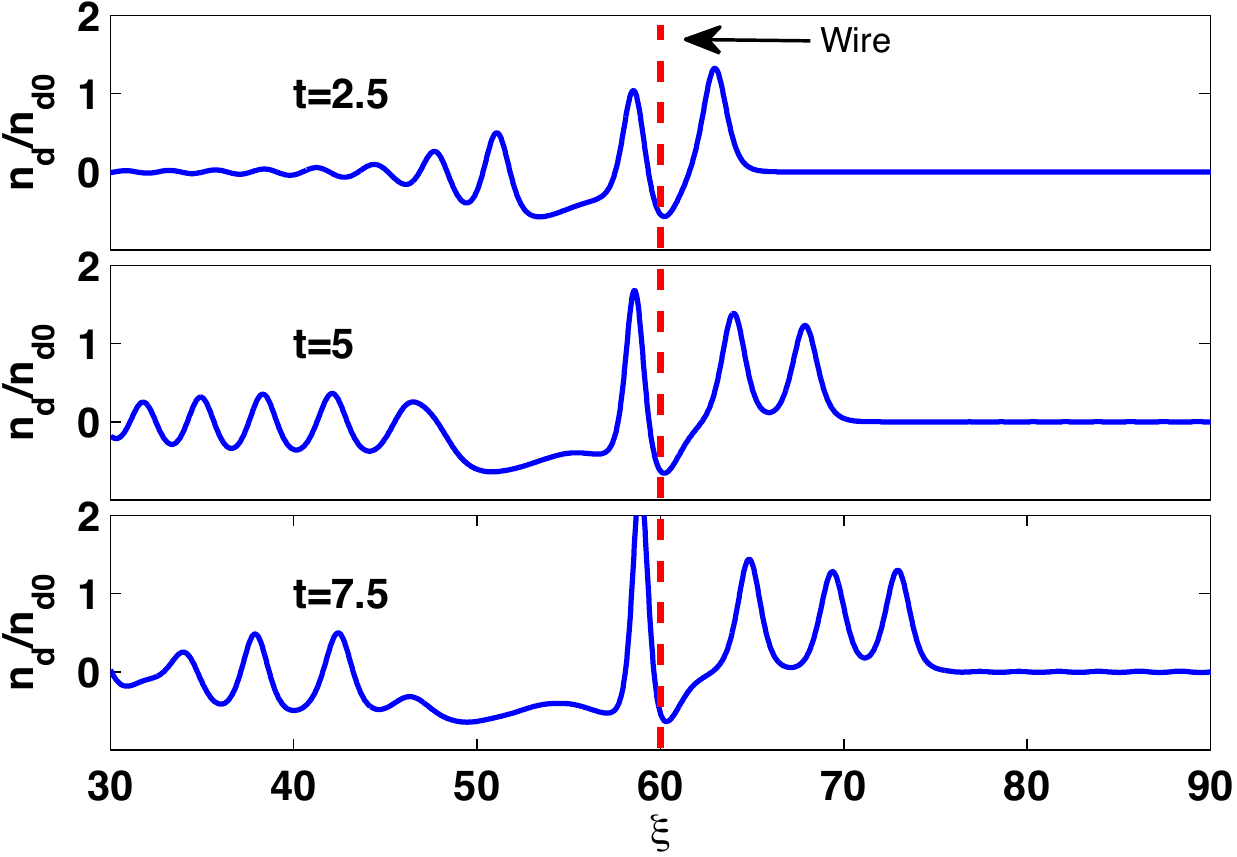}
    \caption{Time evolution of precursor solitons and wakes obtained from a numerical solution of the fKdV equation (\ref{fKdv}). }%
    \label{fig:fig6}%
\end{figure}
Here $\delta = n_{i0}/n_{eo}$, $\sigma_i = T_{i}/T_{e}$ and the equilibrium densities of the electrons, ions and dust component are related by the quasi-neutrality condition $n_{e0} +Z_{d}n_{d0} =n_{i0}$. The term  $S_2(\xi +Ft)$ on the right hand side of (\ref{fKdv}) is the moving charged source term that is moving at a velocity $v_d$ and $F=1 -v_d$.  If the source is held stationary (as in the present experiment) then the fluid (plasma) moves in the opposite direction with speed $v_{d}$. The coefficient $A$ accounts for the change in the low frequency dynamics of the system as compared to the ion acoustic case \cite{Sen2015} due to the involvement of the dust motion. A simple way to recover the form of the forced KdV for the ion acoustic case is to take the limit $\delta = \sigma_{i} =0$ and replace $n_{d1}$ by $n_{i1}$. For our experimental parameters the magnitude of $A$ is approximately $6.2$. In deriving Eq.~1, we have neglected dust neutral collisional damping rate $\nu_{dn}$ of the solitonic solutions which can be calculated by using the standard formula, $\nu_{dn}=\frac{4}{3}\delta\pi a^2 m_n n_n  C_n/m_d$ \cite{Epstein1924}.  For argon gas of pressure $P=0.090$ mbar, $m_n \sim 6.66\times 10^{-26}$ kg, $n_n \sim 1.87\times 10^{21} $m$^{-3}$ and average neutral velocity $C_n \sim 428$ m/sec (assuming the gas temperature is 0.03eV). The Epstein coefficient ($\delta$) is measured in our device as $\delta \sim1.2 $ \cite{Jaiswal2015}.  Using all these parameters $\nu_{dn}$ comes out to be $\approx 9 s^{-1}$.
The soliton energy decays as $e^{-\nu_{dn}t}$ while its amplitude and width scale as $e^{-2\nu_{dn}t/3}$ and $e^{\nu_{dn}t/3}$ respectively \cite{zhdanov2002, Samsonov2002}. For a soliton speed of $v_s \sim 4.15$ cm/sec, the damping length is therefore approximately $3 v_s/2\nu_{dn} \sim 7$ mm which is about nine times larger than the width ($\Delta \sim 0.8 $ mm) of the soliton. Hence to a first approximation one can neglect dissipative contributions \cite{Samsonov2002} to Eq.~1. In our experiment we observe the development of a soliton to take place over an average distance of about $5$ mm and its average propagation length without significant decay is about $9.5$ mm as can be seen in Fig.~4 where the scale measure of 10 mm is clearly marked. Our model equation is therefore physically justified for describing soliton propagation over these distances. Over a longer time scale the calculated  dissipation would lead to an exponential decay of the soliton amplitude and a broadening of its width.
\begin{figure}[!ht]
\includegraphics[scale=0.90]{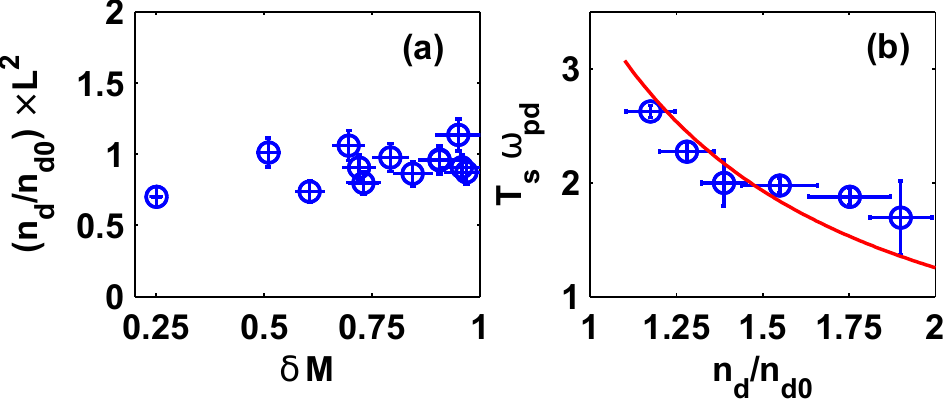}
\caption{\label{fig:fkdv_props} {Variation of (a) soliton parameter (amplitude $\times$ width$^2$) with excess Mach number $\delta M = M -1$ where the Mach number $M$ is normalized to the dust acoustic wave speed and (b) time interval between the generation of two fully developed solitons with the amplitude. The solid line in Fig.~\ref{fig:fkdv_props}(b) is a plot of the curve $T_s\omega_{pd}=\alpha/(n_d/n_{d0})^{3/2}$, where $\alpha=3.54$. }}
\end{figure}
Fig.~\ref{fig:fig6} shows typical time evolution plots of $n_{d1}$ for $v_d>1$ from a numerical solution of eq.(\ref{fKdv}) with a Gaussian source term $S_2$ to model the wire generated potential hill. To compare with the experimental observations the solutions have been plotted in the frame of the moving source which is now stationary at the point marked by the dashed line. As can be seen the source periodically excites solitons ahead of its path that travel away at a faster speed. Weaker excitations consisting of wakes are seen to emerge in the downstream direction. Thus the fKdV model provides a consistent description of our experimental observations. To further confirm the consistency of our experimental findings with the model description we have carried out two more tests. A well known property of a KdV soliton is that the product of its amplitude with the square of its width is a constant quantity \cite{Miura1976}. This property also holds in the weakly dissipative limit where the decrease in the amplitude and broadening of the width due to dissipation are such as to still keep the product constant. In Fig.~\ref{fig:fkdv_props}(a) we have plotted the measured value of this quantity for the solitonic structures of different sizes observed in our experiments. As can be seen this quantity is nearly a constant (of value close to unity) thereby confirming the KdV like solitonic nature of these excitations. The fKdV model also predicts a scaling law for the inter-solitonic intervals, namely, $T_s \propto A^{-3/2}$ where $T_s$ is the interval between generation of the solitons and $A$ is the amplitude of the soliton \cite{Wu1987}. In Fig.~\ref{fig:fkdv_props}(b) we have plotted the experimentally observed time intervals (normalized to the dust plasma frequency) against the soliton amplitudes ($n_{d1}/n_{d0}$) and the time intervals are seen to decrease monotonically with increasing amplitudes of the solitons. The solid curve is an analytic plot of the function $T_s\omega_{pd}=\alpha/(n_d/n_{d0})^{3/2}$ with $\omega_{pd} \approx 30 Hz$, $\alpha=3.54$ and serves as a visual aid to discern the trend in the data points.\par
In conclusion, we have experimentally demonstrated, for the first time, the existence of precursor soliton excitations in a plasma medium caused by a supersonic flow of the plasma over an electrostatic potential hill.  The mechanism underlying the excitation of such solitons can be understood in terms of the
following physical picture. When an object moves through a fluid at a subcritical (subsonic) velocity the build-up of the density perturbation in front of it is able to move away as linear waves traveling at the phase speed of the linear excitations  of the medium. This leads to the formation of wake fields. However when the object speed is supercritical (supersonic) such a reduction of the density perturbation is not possible and as it starts to build up nonlinear effects begin to become important. It is this process that eventually leads to the formation of solitons i.e. when nonlinear effects are balanced by dispersion, and these structures now travel ahead of the moving body.  We would like to emphasize that these solitonic excitations arising due to the interaction of a flowing plasma with a barrier are fundamentally different from past observations of standard dust acoustic solitons that are created by impulsive excitations of the plasma \cite{Nosenko2002,Samsonov2002, Bandyopadhyay2008, Heidemann2009, Zhdanov2010}. They also go beyond the usual wake field excitations seen behind a moving charged object in a plasma and are a distinctly {\it forewake} phenomenon analogous to that observed in hydrodynamic experiments and are associated with transcritical flows.
Transcritical (supersonic) flows of plasmas can occur in many natural situations such as in astrophysical jets, 
solar winds etc. The encounter of such flows with a stationary charged object can recreate the situation discussed in our experiment. Likewise, instances of charged objects moving at supersonic speeds in a plasma 
are those of satellites (which naturally acquire surface charges) orbiting the earth in the ionospheric region, high energy ion beams impinging on targets in inertial fusion schemes etc. It would be interesting to look for precursor solitons in such situations since their presence can have significant practical implications. For example, in particle beam fusion applications they could impact the heating and or compression dynamics of the target pellet. In the satellite orbital regions above the earth, the detection of such solitonic excitations from charged debris objects could help provide an early warning scheme for satellites to avoid collisions with such objects \cite{Sen2015}. The study of such precursor soliton excitations, which surprisingly has not received any attention so far in the plasma physics community, can open up new areas of experimental and theoretical research in the field and our present experimental investigations are a first step in that direction. 
\begin{acknowledgments} 
We are thankful to G. Ganguli and M. Lampe for their insightful remarks and helpful suggestions on our experiment. One of us (AS) gratefully  acknowledges the support provided for this research by Grant No. FA2386-13-1-4077 AOARD 134077.
\end{acknowledgments}

\end{document}